\RequirePackage{fix-cm}
\documentclass[smallextended]{svjour3}       
\smartqed  
\usepackage{graphicx}
\begin{document}

\title{Assisted  coherence distillation of certain mixed  states
}


\author{Xiao-Li Wang  \and Qiu-Ling Yue \and  Su-Juan Qin$^{\dag}$ 
}


\institute{X.-L. Wang         \and Q.-L. Yue
          \and S.-J. Qin$^{\dag}$  \at
              State Key Laboratory of Networking and Switching Technology, Beijing University of Posts and Telecommunications, Beijing 100876, China \\
              $^{\dag}$\email{qsujuan@bupt.edu.cn}           
           \and
           X.-L. Wang \at
              School of Mathematics and Information Science, Henan Polytechnic University, Jiaozuo, 454000, China
}

\date{Received: date / Accepted: date}

\maketitle

\begin{abstract}
In the task of assisted coherence distillation via the set of operations $X$, where $X$ is either local incoherent operations and classical communication (LICC), local quantum-incoherent operations and classical communication (LQICC), separable incoherent operations (SI), or separable quantum-incoherent operations (SQI), two parties, namely Alice and Bob, share many copies of a bipartite joint state. The aim of the process is to  generate the maximal possible coherence on the subsystem of Bob. In this paper, we investigate the assisted coherence distillation of some special  mixed states, the states with vanished  basis-dependent discord and Werner states. We show that all the four sets of operations  are  equivalent for assisted coherence distillation, whenever Alice and Bob share one of those mixed quantum states. Moreover, we prove that the assisted coherence distillation of the former can reach the upper bound, namely QI relative entropy, while that of the latter can not. Meanwhile, we also present a sufficient condition such that the assistance of Alice via the set of operations $X$ can not help Bob improve his distillable coherence, and this condition is that the  state shared by Alice and Bob has vanished  basis-dependent discord.

\end{abstract}

\section{Introduction}
  Quantum coherence, another embodiment of the superposition principle of states, is essential for many distinctive and captivating characteristics of quantum systems [1-3]. The  quantification of of quantum coherence  was recently proposed by Baumgratz \itshape et al. \upshape [4]. Under this framework, many new coherence quantifiers have been presented since then [5-8]. Meanwhile, various properties of quantum coherence have been investigated such as the connections between quantum coherence and quantum correlations [9-14], the distillation of coherence [15-17], the dynamics under noisy evolution of quantum coherence [18,19], among others. The role of coherence in biological system [20], and thermodynamics [21] has also been explored.

  In the framework of coherence given by Baumgratz \itshape et al. \upshape [4], let $\{|i\rangle\}$ be some fixed reference basis (incoherent basis) in the finite dimensional Hilbert space, and a state is said to be incoherent if it is diagonal in this basis, being of the form $\sum_{i}p_{i}|i\rangle\langle i|$. A quantum operation is identified by a set of Kraus operators $\{K_{l}\}$  satisfying $\sum_{l}K^{\dag}_{l}K_{l}=I$. Specially, a quantum operation is called incoherent if it can be written in the form $\Lambda(\rho)=K_{l}\rho K^{\dag}_{l}$, with incoherent Kraus operators $K_{l}$, i.e., $K_{l} |m\rangle \sim |n\rangle$, where $|m\rangle$ and $|n\rangle$ are elements of the fixed reference basis. As a quantifier of quantum coherence, we will use the relative entropy of coherence, initially defined as
$C_{re}(\rho)=\min_{\sigma\in \mathcal{I}}S(\rho\|\sigma),$
 where $S(\rho\|\sigma)=Tr(\rho\log_{2}\rho)-Tr(\rho\log_{2}\sigma)$ is the quantum relative entropy [22] and the minimization is taken over the set of all incoherent states $\mathcal{I}$. Crucially, the relative entropy of coherence can be evaluate exactly:
 $C_{re}(\rho)= S(\mathrm{\Pi}(\rho))-S(\rho)$ , where $S(\rho)=-Tr(\rho \log_{2}(\rho))$ is the von Neumann entropy [22] and $\mathrm{\Pi}(\rho) =\sum _{i} |i\rangle \langle i| \rho |i\rangle \langle i|$ denotes the von Neumann measurement (dephasing operation) of $\rho$ with respect to the fixed reference basis.
 An important progress within the resource theory of coherence is the operational theory of coherence, which is introduced by Winter and Yang [15]. Particularly, they presented the distillable coherence for any  state $\rho$, i.e., $C_{d}(\rho)=C_{re}(\rho)$, where the distillable coherence $C_{d}(\rho)$ is defined as the maximal rate for extracting maximally coherent  single-qubit states, $|\Psi_{2}\rangle=\frac{1}{\sqrt{2}}(|0\rangle+|1\rangle)$, from a given state $\rho$ via incoherence operations.

For a bipartite system $AB$ shared by Alice and Bob, its reference basis is usually assumed to be the tenser product basis of local bases [6], i.e., $\{|ij\rangle^{AB}\}$, where $\{|i\rangle^{A}\}$ and  $\{|j\rangle^{B}\}$  are  the fixed reference bases of subsystems $A$ and $B$, respectively. In the task of assisted coherence distillation [16, 17], two parties, namely Alice and Bob, share many copies of some bipartite state $\rho^{AB}$ in system $AB$ and their goal is to maximize coherence on Bob's subsystem via special operations. The task of assisted coherence distillation via local quantum-incoherent operations and classical communication (LQICC) was firstly introduced in Ref. [16]. Then, Streltsov \itshape et al. \upshape [17] further studied the assisted coherence distillation via the set of operations $X$, where $X$ is either local incoherent operations and classical communication (LICC), LQICC, separable incoherent operations (SI), or separable quantum-incoherent operations (SQI). In these sets of operations, Bob is restricted to just incoherent operations on his subsystem, since they do not allow for local creation of coherence on Bob's side. The corresponding distillable coherence on Bob's side is denoted by $C_{X}^{A|B}$ and can be given as follows [17]:
$$C_{X}^{A|B}(\rho^{AB})=sup\{R:\mathop {lim} \limits_{n\rightarrow\infty}(\mathop{inf}\limits _{\Lambda\in X}\|Tr_{A}[\Lambda[\rho^{\otimes n}]]-\Psi_{2}^{\otimes \lfloor Rn\rfloor}\|)=0\},$$
where $\lfloor x \rfloor$ is the largest integer below or equal to $x$ and $\Psi_{2}=|\Psi_{2}\rangle\langle\Psi_{2}|^{B}$ is a maximally coherent single-qubit state on Bob's side. For any bipartite state $\rho^{AB}$, the distillable coherence $C_{X}^{A|B}(\rho^{AB})$ is upper bounded by the QI relative entropy $C_{re}^{A|B}(\rho^{AB})= \min_{\sigma^{AB} \in \mathcal{QI}}S(\rho^{AB}\|\sigma^{AB})$, where the minimization is taken over the set of all quantum-incoherent states $QI$ [17]. Quite remarkably, both sets of operations SQI and SI lead to the same maximal performance for all states in the task of assisted coherence distillation. Besides, for all pure states and maximally correlated states in the incoherent basis, all the sets of operations we consider are always equivalent in this task and the distillable coherence can reach the upper bound, namely QI relative entropy. However, we do not know whether the sets of operations such as LICC, LQICC and SI (SQI) are equivalent in this task and whether the distillable coherence can reach the upper bound for general mixed states. Therefore, in this paper, we will investigate the assisted coherence distillation of two special classes of  mixed states. The results suggest that there exist states, namely Werner states,  whose distillable coherence  can not reach the upper bound via the set of operations $X$ we consider here.

\section{ Assisted coherence distillation of states with vanished  basis-dependent discord}
In the following we discuss the scenario where the  state shared by Alice and Bob has vanished  basis-dependent discord [23]. In this section, we will frequently refer to a bipartite system $AB$, and without other stated, we use $\{|i\rangle^{A}\}$ and  $\{|j\rangle^{B}\}$  as the fixed reference bases of subsystems $A$ and $B$, respectively. Before we study this work, we recall the condition of states with vanished basis-dependent discord as follows.

\textbf{Lemma 1}. ( Yadin \itshape{et al.} \upshape{[23]})  Let $\mathrm{\Pi}^{B}$ be the von Neumann measurement with respect to the fixed reference basis $\{|j\rangle^{B}\}$ of subsystem $B$. For any given state $\rho^{AB}$ in system $AB$, let $\rho^{A}$ and $\rho^{B}$ be the reduced states of  $\rho^{AB}$ on subsystem $A$ and $B$, respectively.  Then, its basis-dependent discord $D^{A|B}_{\mathrm{\Pi}^{B}}(\rho^{AB})$ is zero, i.e.
$D^{A|B}_{\mathrm{\Pi}^{B}}(\rho^{AB})=S(\rho^{AB}\|\rho^{A}\otimes \rho^{B})-S(\mathrm{\Pi}^{B}(\rho^{AB})\|\rho^{A}\otimes \mathrm{\Pi}^{B}(\rho^{B}))=0$, if and only if
there exists a decomposition of $ \rho_{AB}$,
 \begin{equation}
 \rho^{AB}=\sum_{\alpha}p_{\alpha} \rho^{A}_{\alpha}\bigotimes \rho^{B}_{\alpha},
 \end{equation}
such that  $\rho^{B}_{\alpha}$ are perfectly distinguishable by projective measurements in the fixed reference basis $\{|j\rangle ^{B}\}$.

It is well known that coherence is a basis-dependent measure of quantumness in single systems, and similarly, $D^{A|B}_{\mathrm{\Pi}^{B}}$ can be seen as a basis-dependent measure of quantumness of correlation [23]. From the Lemma 1, we know that the states with vanished basis-dependent discord are  special separable states whose $\rho^{B}_{\alpha}$ have disjoint coherence support.  Another important result we will refer to is the following Lemma 2 which is about the distillable coherence $C_{X}^{A|B}$ via the set of operations $X$ we consider here.

\textbf{Lemma 2}. (Streltsov  \itshape{et al.} \upshape[17])  For an arbitrary bipartite state $\rho=\rho^{AB}$ in system $AB$, the following inequality holds:
\begin{equation}
C_{LICC}^{A|B}(\rho)\leq C_{LQICC}^{A|B}(\rho)\leq C_{SI}^{A|B}(\rho)= C_{SQI}^{A|B}(\rho)\leq C_{re}^{A|B}(\rho),
\end{equation}

Lemma 2 presents the power of all the sets of operations such as LICC, LQICC, SI and SQI in the task of assisted coherence distillation for an arbitrary bipartite state. Equipped with these results, we are now in position to prove the following theorem.

\textbf{Theorem 3}. For any bipartite state $\rho=\rho^{AB}$ given in Eq.$(1)$, the following equality holds:
\begin{equation}
C_{LICC}^{A|B}(\rho)=C_{LQICC}^{A|B}(\rho)= C_{SI}^{A|B}(\rho)= C_{SQI}^{A|B}(\rho)\nonumber
                         = C_{re}^{A|B}(\rho)= C_{re}(\rho^{B}),
\end{equation}
 where $\rho^{B}$ is the reduced state of  $\rho^{AB}$ on subsystem $B$.

\textbf{Proof}. Note that the QI relative entropy of any bipartite state $\rho$ can be evaluated exactly [16]:
\begin{equation}
C_{re}^{A|B}(\rho)=S(\mathrm{\Pi}^{B}(\rho))-S(\rho)=D^{A|B}_{\mathrm{\Pi}^{B}}(\rho)+C_{re}(\rho^{B}).
\end{equation}
Where $\mathrm{\Pi}^{B}$ is the von Neumann measurement in the fixed reference basis $\{|j\rangle^{B}\}$ and $D^{A|B}_{\mathrm{\Pi}^{B}}(\rho)$ is the basis-dependent discord as given in Lemma 1. Combining Eq. $(4)$ and Eq. $(2)$ in Lemma 2, we get the inequality
\begin{equation}
 C_{LICC}^{A|B}(\rho)\leq C_{LQICC}^{A|B}(\rho)\leq C_{SI}^{A|B}(\rho)= C_{SQI}^{A|B}(\rho)\leq D^{A|B}_{\mathrm{\Pi}^{B}}(\rho)+C_{re}(\rho^{B}).
\end{equation}
Besides, it is straightforward that $C_{X}^{A|B}(\rho^{AB})\geq C_{re}(\rho^{B})$ , where $X$ is either LICC, LQICC, SI or SQI.
According to Lemma 1,  for any bipartite state $\rho=\rho^{AB}$ given in Eq.$(1)$, its basis-dependent discord $D^{A|B}_{\mathrm{\Pi}^{B}}(\rho)$ is equal to zero. Applying Eq. $(5)$ and combing the aforementioned results we arrive at the desired  equality given in Eq. $(3)$.  \qed

From Theorem 3, we obtain that for all states with vanished (incoherent) basis-dependent discord, there does not exist basis-dependent quantum correlation between Alice and Bob to help Bob improve his distillable coherence. However, we do not know whether the necessity of Theorem 3 is true. In other words, if the distillable coherence $C_{X}^{A|B}(\rho^{AB})$ via the set of operations $X$, is equal to the relative entropy of coherence $C_{re}(\rho^{B})$, does the state $\rho^{AB}$ has  the form given in Eq. $(1)$? However, as will see in the next section, the distillable coherence of a  Werner state can not reach the upper bound and it means that the basis-dependent discord $D^{A|B}_{\mathrm{\Pi}^{B}}(\rho^{AB})$ of a Werner state can not be completely transformed into the coherence on Bob's side via the set of operations $X$ we consider here.

\section{ Assisted  coherence distillation of Werner states}
Here, we consider assisted coherence of distillation for the Werner states of the form
\begin{equation}
\rho^{AB}=p|\Phi_{+}\rangle \langle\Phi_{+}|+\frac{1-p}{4}I,
\end{equation}
 where $ p\in(0,1)$,  and $|\Phi_{+}\rangle$ is the Bell state denoted as $|\Phi_{+}\rangle=\frac{1}{\sqrt{2}}(|00\rangle+|11\rangle$ (or other Bell states [22]). Here $I$ is the identity operator on the composite system or on the marginal systems, depending on the context. For the two-qubit system,  we choose the computable basis as the fixed reference basis of each subsystem. In this case, the reduced state of Bob $\rho^{B}$ is incoherent and $C_{re}(\rho^{B})=0$. Then, the distillable coherence of Bob via the set of operations $X$ comes from the basis-dependent quantum correlation between Alice and Bob. Moreover, in the following theorem we show that the distillable coherence for a Werner state can not reach the upper bound, namely QI relative entropy  and in this case all sets of operations we consider here are equivalent for assisted coherence distillation.

 \textbf{Theorem 4}. For any  Werner state $\rho=\rho^{AB}$ given in Eq. $(6)$, the following inequality holds:
 \begin{equation}
C_{LICC}^{A|B}(\rho)= C_{LQICC}^{A|B}(\rho)= C_{SI}^{A|B}(\rho)= C_{SQI}^{A|B}(\rho) < C_{re}^{A|B}(\rho),
\end{equation}

 \textbf{Proof}. In the first step of the proof we show that for any Werner state $\rho^{AB}$, its distillable coherence $C_{LQICC}^{A|B}(\rho^{AB})$ is strictly less than its QI relative entropy $C_{re}^{A|B}(\rho^{AB})$. By direct calculations, the QI relative entropy of the Werner state $\rho^{AB}$ is
 \begin{eqnarray}
C_{re}^{A|B}&(\rho^{AB})&= D^{A|B}_{\mathrm{\Pi}^{B}}(\rho^{AB})\nonumber \\
&=&\frac{1-p}{4}\log_{2}(1-p)-\frac{1+p}{2}\log_{2}(1+p)+\frac{1+3p}{4}\log_{2}(1+3p).
\end{eqnarray}
where  $ p\in(0,1)$. Next, let Alice use X basis to measure her subsystem. If the result of Alice's measurement is $1$, the post-measurement state of Bob is
\begin{equation}
\rho^{B}_{+}=p|+\rangle\langle+|+(1-p)\frac{I}{2},
\end{equation}
where $|+\rangle=|\Psi_{2}\rangle=\frac{1}{\sqrt{2}}(|0\rangle+|1\rangle)$ is a maximally coherent single-qubit state, and in this case Bob does not need to do anything. If the result of Alice's measurement is $-1$, Alice informs  Bob her result and Bob performs the incoherent Z gate $|0\rangle\langle0|-|1\rangle\langle1|$ on his subsystem to obtain the state with the form given in Eq. $(9)$. Repeat this process, and Bob obtains many copies of $\rho^{B}_{+}$. Then, the distillable coherence of Bob via this LQICC protocol  is
\begin{equation}
 C_{re}(\rho^{B}_{+})=\frac{1+p}{2}\log_{2}(1+p)+\frac{1-p}{2}\log_{2}(1-p).
\end{equation}

 Consequently, we show the fact that $C_{re}(\rho^{B}_{+})$ is the largest distillable coherence of Bob assisted by Alice through  all  LQICC protocols.
Suppose that Alice performs a local quantum measurement $\mathcal{E}$ on her subsystem, and informs Bob her results. According to the results of Alice, Bob performs the incoherent unitary  operations on his subsystem, and then the state on Bob's side can be denoted as
$\rho^{B}_{\mathcal{E}}= p\tilde{\rho}^{B}_{\mathcal{E}}+(1-p)\frac{I_{B}}{2}.$
With the convexity of the relative entropy of coherence [4], we only need to consider the case that $\tilde{\rho}^{B}_{\mathcal{E}}$ is a pure state, which can be written in the Bloch representation,
$$ \tilde{\rho}^{B}_{\mathcal{E}}=
\frac{1}{2}\left(\begin{array}{cc}1+z & x-iy \\ x+iy & 1-z \end{array}\right),$$
where $x,y$ and $z$ are real and satisfy the equation
\begin{equation}
x^{2}+y^{2}+z^{2}=1.
\end{equation}
In this case,  direct calculations show that
$$ \begin{array}{ccl}
C_{re}(\rho^{B}_{\mathcal{E}})&=&\frac{1-p}{2}\log_{2}(1-p)+\frac{1+p}{2}\log_{2}(1+p) \\ \nonumber
& &-\frac{1+pz}{2}\log_{2}(1+pz)-\frac{1-pz}{2}\log_{2}(1-pz).
 \end{array}$$
Examine the maximum value of $C_{re}(\rho^{B}_{\mathcal{E}})$ under the  constraint condition Eq. $(11)$, and we obtain the maximum value $\mathcal{C}_{re}(\rho^{B}_{+})$ at $x=1$ and $y=z=0$. Then, it follows that $\mathcal{C}_{re}(\rho^{B}_{\mathcal{E}})\leq C_{re}(\rho^{B}_{+})$. Therefore,
 for the Werner state $\rho^{AB}$, it holds that $C_{LQICC}^{A|B}(\rho^{AB})= C_{re}(\rho^{B}_{+})$.

Now, we show that $\mathcal{C}_{re}(\rho^{B}_{+})$ is strictly less than the upper bound $C_{re}^{A|B}(\rho^{AB})$.  Let Eq. $(8)$ subtract Eq. $(10)$ and the difference is denoted as
\begin{eqnarray}
 f(p)&=& C_{re}^{A|B}(\rho^{AB})-C_{re}(\rho^{B}_{+})\nonumber \\
     &=& \frac{1+3p}{4}\log_{2}(1+3p)-\frac{1-p}{4}\log_{2}(1-p)-(1+p)\log_{2}(1+p),
 \end{eqnarray}
with $ p\in(0,1)$. Note that $f(0)=f(1)=0, f(\frac{1}{3})>0$ and $f''(p)=\frac{1-3p}{(1+3p)(1-p^{2})\ln2}$. Then, $f(p)$ is strictly convex for all
 $ p\in(0,\frac{1}{3})$ and strictly concave for all $ p\in(\frac{1}{3},1)$. Besides, using $f'(p)=0$, we get the minimum value at $p=0$.
  Therefore, we obtain that $f(p)>0$ for all $ p\in(0,1)$ and plot the function $f(p)$ in Fig. $1$ and Fig. $2$ , which highlights the distillable coherence $C_{re}(\rho^{B}_{+})$ via the set of operations $X$ is strictly less than the upper bound, namely  QI relative entropy for any Werner state given in Eq. $(6)$.

 \begin{figure}
\small
\centering
\includegraphics[scale=0.4]{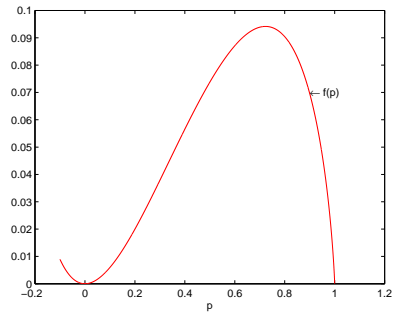}
\caption{ Graph of function $f(p)$ which is the difference between the QI relative entropy and the distillable coherence  via LQICC for any Werner state given in Eq. $(6)$.}
\end{figure}
\begin{figure}
\small
\centering
\includegraphics[scale=0.4]{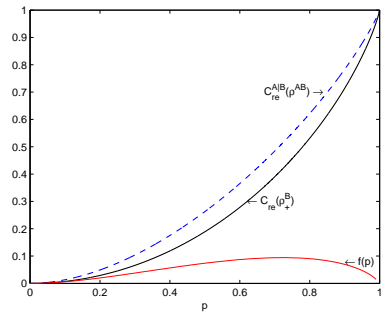}
\caption{ Distillable coherence of Bob is strictly less than the upper bound for any Werner state $\rho^{AB}$ given in Eq. $(6)$ in the task of assisted coherence distillation. We plot the distillable coherence $C_{re}(\rho^{B}_{+})$ (the black line), the upper bound $C_{re}^{A|B}(\rho^{AB})$ (the dashed line), and the function $f(p)$ given in Eq. $(12)$ (the red line).}
\end{figure}

 Finally, we show that for the Werner state $\rho_{AB}$, the distillable coherence of Bob is also equal to  $\mathcal{C}_{re}(\rho^{B}_{+})$ via other sets of operations such as SQI, SI and LICC. Since LQICC is a subset of SQI and SI, the similar approach as above leads to the equality:
$C_{SI}^{A|B}(\rho)= C_{SQI}^{A|B}(\rho)=C_{re}(\rho^{B}_{+}).$  In order to prove that $ C_{LICC}^{A|B}(\rho)=C_{re}(\rho^{B}_{+})$, we present a LICC protocol achieving the rate $C_{re}(\rho^{B}_{+})$. Let Alice perform an erasing (incoherent) measurement on her subsystem [24], whose Kraus operators are
$$ K_{1}^{A}=\left(\begin{array}{cc} \frac{1}{\sqrt{2}}i & \frac{1}{\sqrt{2}} \\ 0 & 0\end{array}\right),
K_{2}^{A}=\left(\begin{array}{cc} -\frac{1}{\sqrt{2}}i & \frac{1}{\sqrt{2}} \\ 0 & 0\end{array}\right),$$
with $(K_{1}^{A})^{\dag}K_{1}^{A}+(K_{2}^{A})^{\dag}K_{2}^{A}=I$. If the result of Alice's measurement is 1, Bob's post-measurement state is $\rho^{B}_{1}=p(\frac{i|0\rangle + |1\rangle}{\sqrt{2}})(\frac{-i\langle0| + \langle1|}{\sqrt{2}})+(1-p)\frac{I}{2}$. In this case,  Bob can get the state $\rho^{B}_{+}$ by applying first an X gate $|1\rangle\langle0|+|0\rangle\langle1|$  and then an incoherent unitary gate $|0\rangle\langle0|-i|1\rangle\langle1|$ to $\rho^{B}_{1}$. If the result of Alice's measurement is 2, Bob's post-measurement state is
$\rho^{B}_{2}=p(\frac{-i|0\rangle + |1\rangle}{\sqrt{2}})(\frac{i\langle0| + \langle1|}{\sqrt{2}})+(1-p)\frac{I}{2}$, and then Bob can get $\rho^{B}_{+}$   by applying first an X gate $|1\rangle\langle0|+|0\rangle\langle1|$  and then a phase gate $|0\rangle\langle0|+ i|1\rangle\langle1|$ to $\rho^{B}_{2}$. Therefore, via LICC, the distillable coherence of Bob is $C_{re}(\rho^{B}_{+})$ with the similar arguments as above. These results imply the inequality given in Eq. $(7)$. \qed

 Note that for standard coherence distillation the relative entropy coherence is in fact equal to the optimal distillation rate [15]. In the task of assisted coherence distillation via the set of operations $X$, the QI relative entropy $C_{re}^{A|B}$ is an upper bound on $C_{X}^{A|B}$ [16, 17]. However, by Theorem 4, we know that the distillation coherence $C_{X}^{A|B}$  is not always able to reach the upper bound, namely QI relative entropy. It is surprising that for a  Werner state $\rho^{AB}$ given in Eq. $(6)$ the distillable coherence $C_{X}^{A|B}(\rho^{AB})$ does not depend on  the particular set of operations performed by Alice and Bob. In particular, in this case the optimal distillation rate can already be obtained by the weakest set of operations LICC. The other sets of operations,  such as LQICC, SQI, and SI, can not provide any advantage to improve the distillable coherence on Bob's side. Through the above results, we strongly conjecture that all the sets of operations we consider here are equivalent in the task of assisted coherence distillation for any mixed state and
we leave this question open for further study.

\section{Conclusions}
In this paper, we have discussed the assisted coherence distillation of some mixed states, such as  the states with vanished (incoherent) basis-dependent discord and Werner states, via the set of operations $X$, where $X$ is either LICC, LQICC, SI or SQI. For these mixed states, all sets of operations we consider here are equivalent for assisted coherence distillation. In particular, we have provided a sufficient condition such that Alice's assistance via the set of operations $X$ can not help Bob improve his distillable coherence, and the condition is that the  state shared by Alice and Bob  has vanished basis-dependent discord.  Moreover, we have proved that the distillable coherence of a Werner state via the set of operations $X$ can not reach the upper bound, namely QI relative entropy. This result suggest that there exist a state for example the Werner state whose distillable coherence is not its QI relative entropy, even through the largest set of operations SQI. We hope that our work helps to well understand the resource theory of coherence in distributed scenarios.

%




\end{document}